\documentclass[preprint, showpacs, aps, nofootinbib]{revtex4}

\usepackage{graphicx}
\usepackage{dcolumn}
\usepackage{bm}

\def\nslash{\rlap{\hspace{0.02cm}/}{n}}

\def\beq{\begin{eqnarray}}
\def\eeq{\end  {eqnarray}}

\def\GeV{{\rm GeV} }

\def\non{\nonumber}

\def\as{\alpha_s}

\def\lqcd{\Lambda_{\rm QCD}}

\newcommand\epjc{Eur.\ Phys.\ J.\ C }

\newcommand\npb{Nucl.\ Phys.\ B }
\newcommand\npps{Nucl.\ Phys.\ B (Proc.\ Suppl.) }
\newcommand\plb{Phys.\ Lett.\ B }
\newcommand\zpc{Z.\ Phys.\ C }

\begin{document}

\title{The QCD factorization in $B \to DKK$ decays}



\author{Zheng-Tao Wei}

\affiliation{
Departamento de F\'{\i}sica Te\'orica, Universidad de Valencia, \\
 E-46100, Burjassot, Valencia, Spain and \\
Institute of Physics, Academia Sinica, Taipei,
 115 Taiwan, ROC}

\begin{abstract}
A study of hadron pair production mechanism is motivated by the
recent observed decays $\bar B^0\to D^{(*)+}K^-K^0$. One novel
phenomenon is threshold enhancement of the kaon pair production. We
show that these decays in the heavy quark mass limit can be
factorized into a generalized form. The new non-perturbative
quantity is the generalized distribution amplitude which describes
how a quark-antiquark pair transmits into the hadron pair. A proof
of factorization of $\bar B^0\to D^{(*)+}K^-K^0$ decays to
all-orders is performed by using the soft-collinear effective
theory. The phenomenological application is discussed in brief.
\end{abstract}

\pacs{13.25.Hw, 12.38.Bx}

\maketitle

\section{Introduction}

Recently, the three body charmful decays $B\to D^{(*)}K^-K^{(*)0}$
were first observed by the BELLE Collaboration \cite{Belle}. They
also found a novel phenomenon that the mass spectra of the
$K^-K^{(*)0}$ pair peaks near their mass threshold. In \cite{CHST},
a factorization approach which extends the conventional (or say,
naive) factorization approach in B-meson two body decays \cite{BSW}
was proposed to study $\bar B^0\to D^{(*)+}K^-K^{0}$ decays. The
transition matrix element of $\bar B^0\to D^{(*)+}K^-K^0$ is
factorized into a $B\to D$ form factor multiplied by a $K^-K^0$ weak
form factor. The $K^-K^0$ form factor is constrained from the
experimental data for the time-like electromagnetic (EM) kaon form
factors. The branching ratios obtained in this approach agree with
the experiment. The phenomenological success indicates that the
generalized factorization assumption may be valid as the leading
approximation of an asymptotic expansion. The purpose of this study
is to establish a theoretical foundation of the factorization for
B-meson multi-body decays, in particular for $B\to DKK$ decays in
QCD.

The factorization approach for the non-leptonic, two body B-meson
decays had been used for a long time \cite{BSW}. The basic idea is
that a hadronic matrix element of the four-quark operator is
factorized into a multiplication of two simpler matrix elements
which one is represented by a form factor and another by a decay
constant. An argument based on ``color transparency" was proposed as
a mechanism of the factorization \cite{Bjorken}. The QCD
factorization method develops the naive factorization into a
rigorous and systematic framework \cite{BBNS}. The main ingredient
is factorization, the separation of perturbative from
non-perturbative dynamics, a standard method in perturbative QCD for
the hard exclusive processes \cite{BL}. The naive factorization
approach is its lowest order (the strong coupling constant $\as$)
approximation. A proof of factorization at two-loop order in $B\to
D\pi$ decays is given by using the methods of momentum regions and
Ward identities. The all-orders proof is done in the recently
developed soft-collinear effective theory \cite{B2}. The proof of
factorization in the soft-collinear effective theory has several
advantages: the explicit gauge invariance at the classical level can
be utilized; the operator language is simpler than the diagrammatic
analysis; the separation of the perturbative from the
non-perturbative dynamics is easily done in effective theory because
they occur at disparate energy scales. We will use the
soft-collinear effective theory to give a proof of factorization for
$\bar B^0\to D^{(*)+}K^-K^0$ decays.

The exclusive, non-leptonic three body B meson decays are usually
more complicate than two body case. However, the hadronic physics of
$\bar B^0\to D^{(*)+}K^-K^0$ decays in the heavy quark limit (the
heavy quark mass $m_b\to\infty$) may be simpler because the final
kaon pair has small invariant mass. It is well-known in QCD that the
large energy behavior of hadron form factor satisfies a dimensional
counting rule \cite{BL}. Consequently, the probability of kaon pair
production at large invariant mass is suppressed by powers of large
energy. This phenomenon is called the from factor suppression. Thus
the dominant contribution comes from the region with small invariant
mass. For the production of kaon pair at small invariant mass, it is
analogous to the production mechanism of two pions in the process
$\gamma^* \gamma\to\pi\pi$ with high virtual photons and small
invariant mass of pion pair \cite{DGPT}. According to this
mechanism, $K^-K^0$ mesons are produced by the hadronization of
$\bar u d$ pair emitted from the W-boson and can be described by a
universal, non-perturbative matrix element which generalizes the
standard description of a hadron in QCD \cite{BL}. The new element
is generalized distribution amplitude(s) which is the crossed
version of of the generalized parton distribution (GPD) of hadron
\cite{Ji}.

In the small invariant mass region, the two kaons are nearly
collinear and energetic. Thus, the argument of ``color transparency"
is applicable. Since the $\bar u d$ pair moves fast, time dilation
effect makes the hadronization of kaon pair cannot occur until the
$\bar u d$ pair moves far away from the remaining system. The
transition $\bar u d\to K^-K^0$ with a small invariant mass is soft
and is described by a generalized distribution amplitude of kaon
pair. For the $B\to D$ transition, the light spectator does not
require a hard interaction because B and D mesons are both heavy.
The energetic and collinear $\bar u d$ pair in a color-singlet
configuration decouples from the soft gluon interactions.  So, the
strong interactions between $K^-K^0$ and $BD$ systems occur at short
distance and can be systematically calculated in perturbative QCD.
The above arguments will lead to a factorization form depicted in
Fig. 1. Up to leading order of $\frac{w}{m_b}$ ($w$ being the kaon
pair invariant mass)\footnote{The mass difference between the heavy
quark and heavy meson is neglected.}, the hadronic matrix element of
a four-quark
operator $Q_i$ in the weak effective Hamiltonian is expressed as %
\beq %
\langle DKK| Q_i|\bar B \rangle
  =F^{BD}(w^2)\int_0^1 dz~H_i(z)\Phi^{KK}(\zeta, z, w^2). %
\eeq %
Here, $F^{BD}(w^2)$ is a $B\to D$ transition form factor at the
momentum transfer $w^2$, and $\Phi^{KK}(z, \zeta, w^2)$ is a
generalized distribution amplitude (GDA) of the kaon pair. $H$
denotes a hard scattering kernel which is perturbatively calculable.
This factorization formulae is a natural generalization of the QCD
factorization in $B\to D\pi$ decays given in \cite{BBNS}. Just as
the $B\to D\pi$ decay, the $B\to DKK$ process provides a clean
environment to test the factorization in the three body B meson
decays.

\begin{figure}
\includegraphics*[width=2.5in]{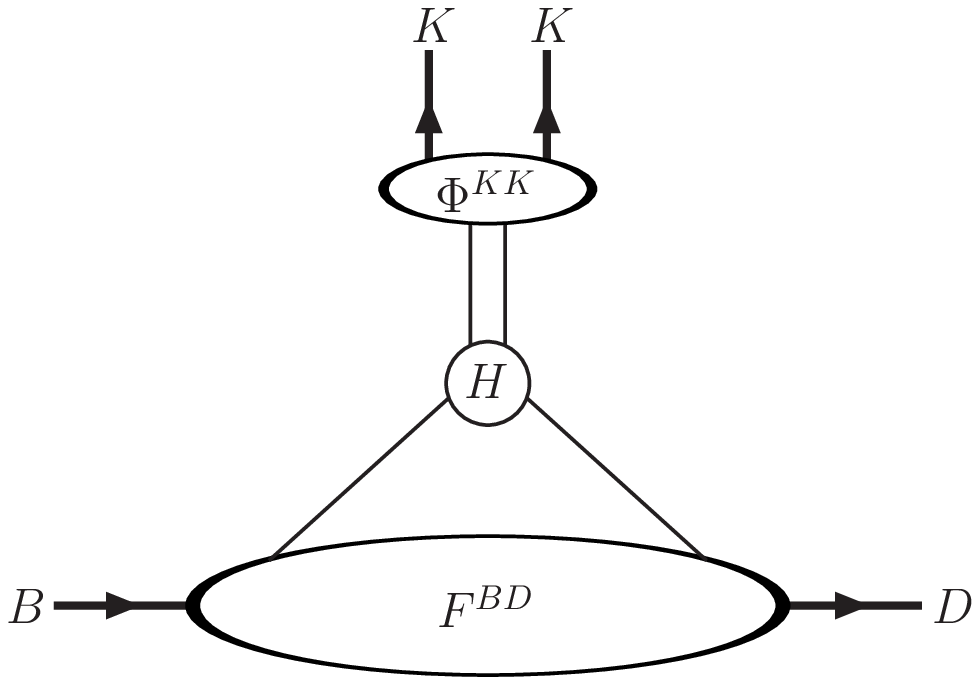}
\caption{Factorization of process $B\to DKK$ in the large $m_B$
 limit.}
\label{fig:BDKK}
\end{figure}

\section{The factorization in $B\to DKK$ decays}

For the sake of illustration, our discussion will concentrate on the
decay $\bar B^0\to D^+K^-K^0$. The extension to $\bar B^0\to
D^{*+}K^-K^0$ decay can easily performed. We shall work in the rest
frame of the B meson. It is convenient to use the light-cone
variables $p^{\mu}=(p^+, p^-, p_{\bot})=n_-\cdot
p\frac{n_+^{\mu}}{2}+ n_+\cdot p\frac{n_-^{\mu}}{2}+p_{\bot}^{\mu}$
with $n_+^{\mu}=(2,0,0_{\bot})$ and $n_-^{\mu}=(0,2,0_{\bot})$ are
two light-like vectors which satisfy $n_+^2=n_-^2=0$, and $n_+\cdot
n_-=2$. Introduce the total momentum of the kaon pair $P=p_1+p_2$
where $p_{1,2}$ are momenta of the $K^-,K^0$ respectively, and
invariant mass $P^2=w^2$. The momentum $P$ is chosen to be mainly in
the ``$-$" direction. Define the momentum fraction variable
$\zeta\equiv
\frac{p_1^-}{P^-}$.  Under the above conventions, we have %
\beq %
&& %
P_B=(m_B, ~m_B,  ~\vec 0_\bot), ~~~~~ %
P_D=(m_B, ~rm_B, ~\vec 0_\bot), ~~~~~ %
P =\bar r(\eta m_B, ~m_B, ~\vec 0_\bot), \non \\
&& %
p_1=\bar r(\bar\zeta\eta m_B, ~\zeta m_B, ~\vec p_{\bot} ), ~~~~~
p_2=\bar r(\zeta\eta m_B, ~\bar\zeta m_B, ~-\vec p_{\bot}).
\eeq %
where $r=\frac{m_D^2}{m_B^2}$, $\eta=\frac{w^2}{(\bar r m_B)^2}$
and $p_{\bot}^2=(\zeta\bar \zeta w^2-m_K^2)/\bar r ^2$. We have
used the ``bar"-notation for any longitudinal momentum fraction
variable $\bar u=1-u$ throughout this paper. One can obtain the
kinematical constraint on the variable $\zeta$: $\zeta\bar
\zeta\ge \frac{m_K^2}{\bar r^2 w^2}$.

We consider the kinematic region %
\beq %
w^2\ll m_B^2. %
\eeq %
This requirement is guaranteed by  the form factor suppression at
large invariant mass. We will give an argument of the suppression
later.

The factorization of $B\to D\pi$ decays in the soft-collinear
effective theory was studied in \cite{B2, B3} by using the hybrid
position-momentum representation. We will use the position space
formulation in \cite{BCDF, BHLN, Wei} to demonstrate factorization.
We also include discussions on the suppression of higher-Fock states
and endpoint contributions based on the power counting of the
soft-collinear effective theory.

The first step of factorization starts from integrating out the
heavy $W$-boson and the hard gluons with virtualities between $m_W$
and a renormalization scale $\mu$. The weak effective Hamiltonian
is obtained as%
\beq %
H_{eff}=\frac{G_F}{\sqrt 2}V_{ud}^*V_{cb}\left[ C_0(\mu)
Q_0+C_8(\mu) Q_8 \right].
\eeq %
where the four-quark operators are %
\beq %
Q_0=\bar c\gamma^{\mu}(1-\gamma_5)b~\bar d\gamma_{\mu}(1-\gamma_5)u,
~~~~ Q_8=\bar c\gamma^{\mu}(1-\gamma_5)T^A b~\bar
d\gamma_{\mu}(1-\gamma_5)T^A u,
\eeq %
The Wilson coefficients $C_0, C_8$ are given at scale $\mu\sim m_b$.

The low energy effective field theories relevant to our process are
heavy quark effective theory (HQET) and soft-collinear effective
theory. The field degrees of freedom are: the heavy quark fields
$h_{v'}^c$, $h_{v}^b$; the collinear quark fields $\xi_u$,
$\xi_{d}$; the collinear gluon field $A_c$; the soft quark and gluon
fields $q_s$, $A_s$. For collinear momentum, the off-shellness is
$p_c^2\sim w^2$ rather than $\lqcd^2$. The small expansion parameter
should be $\lambda=\frac{w}{m_b}$ (For the $c$ quark mass, we assume
that it is at the same order of $b$ quark mass $m_c\sim m_b$). The
Wilson lines are indispensable elements in HQET and SCET. The soft
Wilson line $W_s$ and collinear Wilson line $W_c$ are very useful to
give a gauge-invariant quantities. For example, $W_c^{\dagger}\xi$
are invariant under the collinear gauge transformations and
$W_s^{\dagger} q_s$ are gauge invariant under the soft gauge
transformations. The definition of  Wilson lines and the power
counting for the SCET fields are given in \cite{B1, BCDF}.

The coupling of soft gluons to collinear quark is a bit complicated.
In \cite{B3}, the authors introduce auxiliary fields and integrate
out the off-shell modes ($\lambda, 1, \lambda)$. The soft Wilson
line is obtained at the current level. In \cite{Wei}, an approach
which is similar to the HQET is used to decouple the soft gluon from
the collinear quark. The two approaches give the same results in
leading order of $\lambda$. We adopt the approach in \cite{Wei} that
the collinear quark field transforms as $\xi\to W_s(n_-) \xi$ under
the soft gluon interactions. For heavy quark, $h_v\to W_s(v)h_v$.
The $W_s(n)$ represents that the soft gluons go along the $n$
direction. After these transformations, the collinear quark and the
heavy quark does not interact with soft gluons.

The next step is to integrate out the hard mode at order of $m_b$
scale. The coupling of collinear gluon to the heavy quark leads to
off-shellness of order $m_b$ which needs to be integrated from the
effective theory. The tree level matching gives collinear Wilson
lines which can be represented in a gauge-invariant form as
$W_c^{\dagger}\xi$. At the loop level matching, the factorization
formulae will be non-local in position space because the collinear
momentum component $p_c^-$ is of the same order as the hard loop
momentum. At leading order of $\lambda$, the four-quark operators
are matched onto the gauge-invariant operators below %
\beq \label{eq:fp}%
Q_{k}=\sum_i\int ds ~\tilde{C}_{ki}(sm_b,\mu,\mu_F)\bar h_{v'}^c
  W_s^{\dagger}(v')\Gamma_i'^{\mu}T^{k'} W_s(v)h_v^b  \non \\
\times \left[\bar\xi_dW_c\right](sn_+)W_s^{\dagger}(n_-)
  \Gamma_{i\mu}T^{k'} W_s(n_-)\left[W_c^{\dagger}\xi_u\right](0).
\eeq %
where $k, k'$ represent $0$ or $8$ and $T^{k'}=1$ for k'=0,
$T^{k'}=T^A$ for $k'=8$. Note that $k$ and $k'$ can be different
because the color-singlet and color-octet currents mix each other by
hard gluon exchange. The $\tilde{C}_{ki}(sm_b,z,\mu,\mu_F)$ are
position space Wilson coefficients. The $\mu$-dependence of
$\tilde{C}_{ki}(sm_b,z,\mu,\mu_F)$ cancels the dependence of $C_{k}$
on the renormalization scale $\mu$. The $\mu_F$ is the factorization
scale which separates the hard modes from the matrix element of the
kaon pair. The appearance of $\mu_F$ starts from two-loop order. In
concept, $\mu_F$ is different from $\mu$. Compared to the relevant
operators given in \cite{B3}, our formula contains explicit soft
Wilson line $W_s(v), W_s(v')$. It is found that hard-soft momentum
region contribution does not cancel at two-loop order \cite{BBNS}
and can be absorbed in the definition of the $B\to D$ transition
form factor.

For $\bar B^0\to D^+\pi^-$ decays, the next step is to prove that
the soft gluons which attach the collinear quarks decouple and then
cancel. This cancelation occurs for both the non-factorizable and
factorizable diagrams for soft gluons. Here, the non-factorizable
diagrams represents the graphs which attach $\bar u, d$ quarks to
$b,c $ quarks. For $\bar B^0\to D^+K^-K^0$, the case is different.
The non-factorizbale soft gluons decouple from the collinear fields
and vanish due to unitarity of the soft Wilson line
$W_s^{\dagger}(n_-)W_s(n_-)=1$ for color-singlet current operator.
But, the factorizable soft gluons does not cancel completely because
the kaon pair system contains the valence $s\bar s$ quarks as well
as a lot of see quarks and gluons with soft momenta of order of $w$.
The non-cancelation of soft dynamics is similar to $B\to D$
transitions where the interactions with the spectator quark in B
meson are soft dominant. So the QCD dynamics of kaon pair is not
collinear dominant as the single kaon or pion. However, the
non-cancelation of soft interactions does not break down the
factorization because they occur far away from the hard interaction
point. The collinear fields in Eq. (\ref{eq:fp}) does not receive
the non-factorizable soft gluon contributions. So, the factorization
formula of Eq. (\ref{eq:fp}) have factorized the hard interaction
with virtualities of order of $m_b$ and the soft interaction with
virtualities of order of $w$.

Because the color conservation of QCD, a color-octet current can not
couple to two color-singlet kaons. Although some literatures discuss
the possibility of color-octet contribution for heavy quarkonium,
there is no indication to introduce the color-octet mechanism for
light hadrons. So the color-octet operator part in Eq. (\ref{eq:fp})
can be set to zero. From the above discussions, we can separate the
hadronic matrix element $\langle D^+K^-K^0| Q_k|\bar B^0\rangle$
into two separate parts
as %
\beq \label{eq:fp2}%
\langle D^+K^-K^0|Q_k|\bar B^0\rangle=\int ds
  ~\tilde{C}_{k}(sm_b,z,\mu,\mu_F)\langle D^+|\bar h_{v'}^c
  W_s^{\dagger}(v')\Gamma'^{\mu} W_s(v)h_v^b |\bar B^0\rangle \non \\
\times \langle K^-K^0|\left[\bar\xi_dW_c\right](sn_+)
  \Gamma_{\mu}\left[W_c^{\dagger}\xi_u\right](0)|0\rangle,~~~~~~
\eeq %

Since $\xi=\frac{\nslash_-\nslash_+}{4}\psi_c$, the Dirac spin
structure in Eq. (\ref{eq:fp2}) has only one choice
$\Gamma=\nslash_+$ in leading order of $\lambda$. The kaon meson is
spin zero, so the axial part of the collinear current operator does
not contribute owing to parity conservation. The spin matrix
$\nslash_+\gamma_{\bot}^{\mu}$ contribution is possible but it does
not appear in leading order of $\lambda$. Consequently, only the
quark vector current is left which is different from the case of a
single kaon. Similarly, for heavy quark Dirac matrix,
$\Gamma'=\nslash_-(1-\gamma_5)$ in our special case.

The matrix element for $B\to D$ transition is defined by %
\beq \label{eq:form}%
\langle D^+(v')|\bar h_{v'}^c W_s^{\dagger}(v')\gamma_{\mu}
 W_s(v)h_v^b |\bar
 B^0(v)\rangle=\sqrt{m_Bm_D}\xi(v\cdot v')(v+v')_{\mu},
\eeq %
The relation between the Isgur-Wise function and the $B\to D$
transition form factor is
$F_+^{BD}=\frac{(m_B+m_D)}{2\sqrt{m_Bm_D}}\xi(v\cdot v')$. The
leading power generalized light-cone distribution amplitude for the
$K^-K^0$ pair is defined by the following matrix
element \cite{DGPT}: %
\beq \label{eq:GDA}%
\langle K^-(p_1)K^0(p_2)|\left[\bar\xi_dW_c\right](sn_+)
  \gamma_{\mu}\left[W_c^{\dagger}\xi_u\right](0) |0\rangle =
  P_{\mu}\int_0^1 dz e^{izP\cdot (sn_+)}\Phi^{K^-K^0}(z, \zeta, w^2).
\eeq %
where $\gamma_{\mu}$ is dominated by $\nslash_+\frac{n_-^{\mu}}{2}$.
Introducing the Wilson coefficient in momentum space as %
\beq \label{eq:ft}  %
C_k(z, \mu/m_b, \mu_F/m_b))=\int ds ~e^{izP\cdot
 (sn_+)}\tilde{C}_{k}(sm_b,z,\mu,\mu_F),
\eeq %

Inserting Eqs. (\ref{eq:form}, \ref{eq:GDA}, \ref{eq:ft}) into Eq.
(\ref{eq:fp2}), we obtain the final factorization form %
\beq \label{eq:finalf}%
\langle D^+K^-K^0|Q_k|\bar B^0\rangle= N~ \xi(v\cdot v', \mu_0)
 \int_0^1 C_k(z, \mu/m_b, \mu_0/\mu_F, \mu_F/m_b)
 \Phi^{K^-K^0}(z, \zeta, w^2, \mu_F).
\eeq %
where $N=\sqrt{m_Bm_D}(v+v')\cdot P$. Here, we have chosen $C_k(z,
\mu/m_b, \mu_0/\mu_F, \mu_F/m_b)$ as a dimensionless functions. In
Eq. (\ref{eq:finalf}), we added a scale $\mu_0$ to show explicitly
that the Isgur-Wise function $\xi(v\cdot v', \mu_0)$ has a different
evolution equation from $\Phi^{K^-K^0}(z, \zeta, w^2, \mu_F)$. This
point is not addressed in the previous literatures.

The Wilson coefficients $C_k(z, \mu/m_b, \mu_0/\mu_F, \mu_F/m_b)$
are infrared finite because they are obtained by matching from the
full theory onto the low energy effective theory. They do not depend
on the details of the low energy dynamics. The convolution form of
the factorized form is due to that both the hard coefficient
function and the generalized distribution amplitude depends on the
light-cone momentum fraction $z$.

The generalized distribution amplitude $\Phi(z, \zeta, w^2)$
provides an important theoretical tool to study the production of
two hadron pair. It is the time-like version of a generalized parton
distribution (GPD) of hadron. Since the study of GDA in B decays is
only at the start, we display some general properties of GDA which
is helpful to understand it.

GDA contains much fruitful physical information. $\Phi(z, \zeta,
w^2)$ depends on three variables: quark fraction $z\equiv
\frac{p_{\bar u}^-}{P^-}$, which describes how the current quark
shares the total momentum; hadron fraction
$\zeta=\frac{p_1^-}{P^-}$, which characterizes the momentum
distribution between two hadrons; and the invariant mass $w^2$. One
special feature of GDA is that it is complex in general. The
imaginary part of $\Phi$ is due to rescattering effects or resonance
contributions. The strong phase shift induced by this soft mechanism
is neither power nor perturbative ($\alpha_s$) suppressed because
the final state interactions between two hadrons occur at low
energies. Thus it gives an origin of a large strong phase. In this
paper, we will not explore this point further since only the
absolute value of $\Phi$ is relevant. Another feature of GDA is that
it does not select all the valence quarks of the kaon pair in the
hard scattering at the quark level. The additional valence quark
pair $\bar s s$ contained in GDA plays no role in the hard
scattering.

The $K^-K^0$ GDA has only one isospin state, i.e. $I=1$. For
iso-vector amplitude $\Phi^{K^-K^0}(z, \zeta, w^2)$, the charge
conjugation invariance gives %
\beq %
\Phi(z, \zeta, w^2)=\Phi(1-z, \zeta, w^2)=-\Phi(z, 1-\zeta, w^2).
\eeq %
The amplitude is odd under $\zeta \longleftrightarrow 1-\zeta$, so
the skewness of the hadron momentum distribution is described by
the $\zeta$ dependence. The GDA $\Phi^{K^-K^0}(z, \zeta, w^2)$ is
normalized as: %
\beq \label{eq:normal} %
\int_0^1 dz~ \Phi^{K^-K^0}(z, \zeta, w^2)=
  (2\zeta-1)F^{K^-K^0}(w^2),
\eeq %
where $F^{K^-K^0}(w^2)$ is the $K^-K^0$ from factor in the
time-like region, which is defined by %
\beq \label{eq:ff}%
\langle K^-(p_1)K^0(p_2)|\bar d(0)\gamma_{\mu} u(0)|0 \rangle
 =(p_1-p_2)_{\mu}F^{K^-K^0}(w^2),
\eeq %
The time-like form factor $F^{K^-K^0}(w^2)$ needs to be determined
from experiment. Eq.(\ref{eq:normal}) means that the time-like form
factor can be interpreted from a more general concept, namely, GDA.

The GDA $\Phi(z, \zeta, w^2, \mu)$ will depend on the
renormalization scale which is at the order of $m_b$ in our case.
Since the scale dependence is only related to the non-local product
of quark fields, the evolution of $\Phi(z, \zeta, w^2, \mu)$ is the
same as the BLER evolution of the pion distribution amplitude
\cite{BLER}. In the limit $\mu\to \infty$, the GDA $\Phi^{K^-K^0}(z,
\zeta, w^2, \mu)$
has the asymptotic form \cite{Polyakov} %
\beq %
\Phi^{K^-K^0}(z, \zeta, w^2)=6z(1-z)(2\zeta -1)F^{K^-K^0}(w^2).
\eeq %
Thus the shape of the kaon pair invariant mass spectrum in $\bar
B^0\to D^+K^-K^0$ in the heavy quark limit is completely
determined by the time-like weak form factor $F^{K^-K^0}(w^2)$.

In the above derivation of the factorization, we have assumed that
the momenta of $\bar u d$ quark pair are collinear and the kaon pair
are dominated by the small invariant mass region. Now, we argue that
they are leading power contributions. The endpoint region which one
quark of $\bar u d$ pair contains most energy while another is soft
is suppressed by its phase space $dz\sim \lambda$. The hard
coefficients contain only logarithmic dependence of $z$ and does not
add more power of $1/z$. For the transition pion form factor, the
consistency of factorization requires the suppression of pion
distribution amplitude at endpoint. The contribution from the large
invariant mass of kaon pair is suppressed by $1/Q^2$ with $Q$ the
large invariant mass. This is the dimensional counting rule for
hadron form factor at large $Q$ \cite{BL}. Here, we derive it from
the SCET power counting.

The power counting is: collinear quark field $\xi\sim \lambda$, kaon
meson state $|K\rangle\sim \lambda^{-1}$. The time-like kaon form
factor is described by the matrix element of $\langle
0|~Q'~|K^-K^0\rangle$ at large $Q$ where the effective operator $Q'$
contains four collinear quarks associated with collinear Wilson
lines and Dirac matrix elements in leading power. The more fields
involved, the higher suppression occurs. We does not investigate the
accurate form for the operator $Q'$.
The scaling for the $\langle 0|~Q'~|K^-K^0\rangle$ is %
\beq %
\langle 0|~Q'~|K^-K^0\rangle\sim \lambda^4 \lambda^{-2}=\lambda^2.
\eeq %
Combining with Eq. (\ref{eq:ff}), we derive the well-known result
$F^{K^-K^0}(Q^2)\sim \lqcd^2/Q^2$. This means that the form factor
$F^{K^-K^0}(w^2)$ at large invariant mass $w^2$ is proportional to
$1/w^2$ and suppressed. This suppression is called the form factor
suppression. One important phenomenon related to the form factor
suppression is that the spectrum of kaon invariant mass is enhanced
by its threshold region. So, the threshold enhancement of mass
spectrum is crucial for the consistency of our factorization method.

The higher Fock-states contribute the subleading power corrections.
In \cite{BBNS}, the authors use explicit calculations to show that
3-particle Fock-state is suppressed by powers of $1/m_b$ in the
heavy quark limit. This conclusion can be proved generally from the
simple dimensional analysis. The $m_b$ scalings for the fields and
meson state are: quark field $\xi\sim m_b^{3/2}$, gluon field
$A_{\mu}\sim m_b^1$ and meson state $|K\rangle\sim m_b^{-1}$. The
scaling dimension of field is determined by its ordinary dimension.
For kaon meson, we use the normalization $\langle
K(p')|K(p)\rangle=2p^0\delta^3({\bf p}-{\bf p'})$ to determine its
dimension. Adding one gluon field to the effective operators means
the increase of dimension by $1$, the hard coefficients must be
suppressed by a power of $m_b$ in order to match the dimension of
$m_b$. Specifically, we consider the gluon field $A_{\nu}=\int_0^1
dv~vx^{\mu}G_{\mu\nu} $ insertion. The scaling for $A_{\nu}$ is
$A_{\nu}\sim m_b^2$, so its contribution is $1/m_b^{2}$ suppressed
compared to the leading term. This result is consistent with the
calculation in \cite{BBNS}. Note that the above dimensional analysis
is only applicable for hard interaction.

Our approach can be considered as the application of the QCD
factorization to three-body B decays. We call our method as QCD
factorization approach. Though it seems that our factorization
approach is substantially different from the standard pQCD framework
in \cite{BL}, their basic ideas are similar. Both are based on the
factorization theorem, which separates short-distance from
long-distance physics in a simple and systematic way. The SCET
simplifies the proof of factorization.

\section{Phenomenological application and discussions}

Now we discuss the phenomenological application of QCD factorization
approach into $\bar B^0\to D^{(*)+}K^-K^0$ decays. The leading
contribution comes from the collinear region where the momenta of
$\bar ud$ quarks and two kaon are replaced by their largest minus
variables, i.e. the invariant mass $w^2$ and transverse momentum are
neglected, %
\beq %
p_1=\bar r(0, ~\zeta \frac{m_B}{\sqrt 2}, ~0_{\bot} ), ~~~~~
p_2=\bar r(0, ~\bar\zeta \frac{m_B}{\sqrt 2}, ~0_{\bot} ) \non \\
p_{\bar u}=\bar r(0, ~z \frac{m_B}{\sqrt 2}, ~0_{\bot} ), ~~~~~
p_{\bar u}=\bar r(0, ~\bar z \frac{m_B}{\sqrt 2}, ~0_{\bot} ).
\eeq %
The validity of the above collinear approximation needs to be
checked by consistency of the perturbative result. To the leading
power of $w/m_B$, the decay amplitudes for
$\bar B^0\to D^{(*)+}K^-K^0$ are %
\beq %
A(\bar B^0\to D^+K^-K^0)=\frac{G_F}{\sqrt 2}V_{ud}^*V_{cb}~
  a_1 F_+^{BD}(w^2)~F^{K^-K^0}(w^2)(2\zeta-1)(m_B^2-m_D^2),
\eeq %
and %
\beq %
A(\bar B^0\to D^{*+}K^-K^0)=\frac{G_F}{\sqrt 2}
  V_{ud}^*V_{cb}~a_1 A_0^{BD}(w^2)~ F^{K^-K^0}(w^2)(2\zeta-1)
  2m_{D^*}\epsilon^*_{D^*}\cdot P,
\eeq %
where $V_{ij}$ are CKM matrix elements, $F_+^{BD}, A_0^{BD}$ are
$B\to D^{(*)}$ transition from factors defined in \cite{BBNS}, $a_1$
is the Wilson coefficient, and $\epsilon^*_{D^*}$ is the
polarization vector of $D^*$. In the above equations, we have used
the asymptotic form for $K^-K^0$ generalized distribution amplitude.

\begin{figure}
\includegraphics*[width=2.6in]{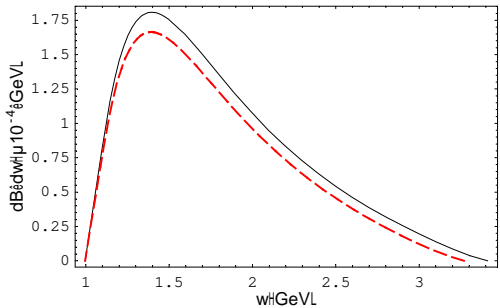}
\caption{The kaon pair invariant mass spectra of $\bar B^0\to
D^+K^-K^0$ (solid line) and $\bar B^0\to D^{*+}K^-K^0$ (dashed
line). \label{fig:spectra}}
\end{figure}

One potential improvement of the QCD factorization approach is that
we can calculate the matrix elements beyond the tree level. We use
the one-loop results for the Wilson coefficient $a_1$ given in
\cite{BBNS}. The theoretical input parameters are chosen as:
$a_1=1.05$, $F_+^{BD}(0)=A_0^{BD}(0)=0.6$. These parameters provide
a well fit to the decay modes of $\bar B{^0}\to D^{(*)+}\pi^-$. The
only unknown input is the $K^-K^0$ weak form factor
$F^{K^-K^0}(w^2)$. This from factor has been constrained via an
isospin relation from the experimental data of time-like
electromagnetic kaon form factors in \cite{CHST}. The obtained
$F^{K^-K^0}(w^2)$ can be approximated as a power-law distribution:
$|F^{K^-K^0}(w^2)|\approx \frac{1.4}{w^2}$. The discrepancy between
this simple model and the best fit result in \cite{CHST} is within
20\% level. It should be noted that this determination method is not
a direct way to extract the $K^-K^0$ weak vector current form
factor. So there are still large theoretical uncertainties coming
from the form factor. Table~\ref{tab:Br} and Fig.~\ref{fig:spectra}
give the numerical results for branching ratios and kaon pair
invariant mass spectra of $\bar B^0\to D^{(*)+}K^-K^0$ decays.

\begin{table}
\caption{ \label{tab:Br} The branching ratios of $\bar B^0\to
 D^{(*)+}K^-K^0$ in units of $10^{-4}$. ``QCD" represents
 the QCD factorization approach.}
\begin{ruledtabular}
\begin{tabular}{lcr}
    & QCD  &  Experiment       \\ \hline
$\bar B^0\to D^+K^-K^0$
   & 1.99 & $1.6\pm 0.8\pm 0.3$ \\ \hline
$\bar B^0\to D^{*+}K^-K^0$
   & 1.77  & $2.0\pm 1.5\pm 0.4$ \\
\end{tabular}
\end{ruledtabular}
\end{table}

Another test of pQCD comes from the ratio of decay rates
\beq %
r_{K^-K^0} \equiv \frac{\Gamma(\bar B^0\to
 D^{*+}K^-K^0)}{\Gamma(\bar B^0\to D^+K^-K^0)}
\approx \Bigg ( \frac{A_0(w^2)}{F_+(w^2)}
\frac{(m_B^2-m_{D^*}^2)}{(m_B^2-m_D^2)} \Bigg )^2 =0.95.
\eeq %
In the above relation, we have neglected the effects caused by the
phase space difference. The use of this ratio can reduce the
theoretical errors caused by the uncertainties of $F^{K^-K^0}(w^2)$.
In QCD factorization approach, the ratio $r_{K^-K^0}$ is slightly
smaller than 1, while in factorization approach it is about 2
\cite{CHST}. The difference between the predictions in the two
approaches lies in the collinear approximation adopted in QCD
factorization approach. Whether this approximation is reasonable or
not is crucial for the validity of applying factorization at the
realistic energy of $m_B$.

From Table~\ref{tab:Br}, the theoretical predictions of branching
ratios are consistent with the experimental data. For the ratio
$r_{K^-K^0}$, it needs further tests. The momentum spectra plotted
in Fig.~\ref{fig:spectra} shows a similar momentum distribution
for pseudoscalar and vector D mesons. The fraction comes from the
range $w<1.5\GeV$ is about 35\% which is not sufficient to
guarantee the validity of the collinear approximation. Note that
this numerical result is based on our insufficient information of
the $K^-K^0$  weak vector current form factor. From physical
considerations, the $K^-K^0$ form factor in the small invariant
mass region is likely to be enhanced by the resonance contribution
or soft re-scattering effects. The expected momentum spectra
should be more concentrated in the mass region close to the
threshold where the two kaons are collinear. This conjecture is
reinforced by the experimental measurement that the fraction of
$B^-\to D^0 K^- K^0$ signal events in the invariant mass range
$w<1.3\GeV$ is 55\%. The best fit $K^-K^0$ form factor in
\cite{CHST} does not satisfy this criterion. If the future
experiment observes that most of the contribution comes from the
small invariant mass region, such as $w<1.5\GeV$, it will provide
a strong support of our conjecture. Furthermore, we suggest to
extract the $K^-K^0$ weak form factor directly from the momentum
spectrum of $\bar B^0\to D^+K^-K^0$.

One conclusion can be obtained from the QCD factorization is that
the power correction is proportional to $w/m_b$. So, the
factorization is only applicable for the invariant mass smaller than
2$\GeV$. For three body baryonic B decays, the threshold energy of
two baryons is higher than 2$\GeV$. So, three body baryonic B decays
is at the margin of the factorization method. Although the $K^-K^0$
form factor is not known accurately at present, we give a crude
estimate about the theoretical errors in $\bar B^0\to
D^{(*)+}K^-K^0$ decays based on power counting. The next-to-leading
power correction at the amplitude level is proportional to
$\frac{2m_K}{m_B}$, which is about $20\%$.  At the decay rate level,
the theoretical accuracy within $40\%$ is possible to be accessible
in QCD factorization method. This accuracy is not as good as that in
$B\to D\pi$, but it is still important in explaining the
experimental data and understanding the hadronic physics of three
body decays.

The principle that the hadron pair produced through quark-antiquark
pair can be applied to other three body B-meson decays, such as
$D\pi\pi$, $\pi\pi K$ etc. For these processes, more generalized
distribution amplitudes are required. Some studies of using GDAs (or
called two-hadron distribution amplitudes) in B decays have been
considered in \cite{MCL}. The detailed exploration of this subject
is also required.

We would like to thank H. Li and C. Chua for many valuable
discussions. This work is partly supported by Grant FPA/2002-0612 of
the Spanish Ministry of Science and Technology,  National Science
Council of R.O.C. under Grant No. NSC 91-2816-M-001-0012-6.

\end{document}